\documentclass[10pt,twocolumn,letterpaper]{article}

\usepackage{iccv}
\usepackage{times}
\usepackage{epsfig}
\usepackage{graphicx}
\usepackage{amsmath}
\usepackage{amssymb}
\usepackage{multirow}
\usepackage{hhline}
\usepackage{bm}
\usepackage{booktabs}
\usepackage{caption}
\usepackage{subcaption}
\usepackage{color}

\usepackage{caption}
\usepackage[pagebackref=true,breaklinks=true,letterpaper=true,colorlinks,bookmarks=false]{hyperref}

\iccvfinalcopy 

\def\iccvPaperID{35} 

\ificcvfinal\fi

\begin{document}

\title{HighEr-Resolution Network for Image Demosaicing and Enhancing}

\author{Kangfu Mei$^{1}$, Juncheng Li$^{2}$, Jiajie Zhang$^{3}$, Haoyu Wu$^{1}$, Jie Li$^{1}$,
Rui Huang$^{14\thanks{Corresponding author: Rui Huang (email: ruihuang@cuhk.edu.cn).
This work is partially supported by funding from Shenzhen Institute of Artificial Intelligence and Robotics for Society, and Robotics Discipline Development Fund from Shenzhen Municipal Government (Grant No. 2016-1418).}}$\\
\normalsize $^{1}$The Chinese University of Hong Kong, Shenzhen~~$^{2}$East China Normal University\\
\normalsize $^{3}$Kuaishou Technology~~$^{4}$Shenzhen Institute of Artificial Intelligence and Robotics for Society\\
{\tt\normalsize \{kanfumei,haoyuwu,jieli1\}@link.cuhk.edu.cn}\\
{\tt\normalsize 51164500049@stu.ecnu.edu.cn~~zhangjiajie@kuaishou.com~~ruihuang@cuhk.edu.cn}
}

\maketitle
\ificcvfinal\thispagestyle{empty}\fi

\begin{abstract}
Neural-networks based image restoration methods tend to use low-resolution image patches for training.
Although higher-resolution image patches can provide more global information, state-of-the-art methods cannot utilize them due to their huge GPU memory usage, as well as the instable training process.
However, plenty of studies have shown that global information is crucial for image restoration tasks like image demosaicing and enhancing.
In this work, we propose a HighEr-Resolution Network (HERN) to fully learning global information in high-resolution image patches. 
To achieve this, the HERN employs two parallel paths to learn image features in two different resolution, respectively.
By combining global-aware features and multi-scale features, our HERN is able to learn global information with feasible GPU memory usage.
Besides, we introduce a progressive training method to solve the instability issue and accelerate model convergence.
On the task of image demosaicing and enhancing, our HERN achieves state-of-the-art performance on the AIM2019 RAW to RGB mapping challenge.
The source code of our implementation is available at \url{https://github.com/MKFMIKU/RAW2RGBNet}.
\end{abstract}

\section{Introduction}
Image enhancement is a hot topic in the field of low-level computer vision, which aims to improve the visual quality of degraded images by using prior-based methods or example-based methods.
Inspired by the powerful non-linear learning ability of neural networks, current methods~\cite{dong2015image, ignatov2017dslr} tend to use deep neural networks to learn the mapping between the degraded RGB image with the desired RGB target.
Theses methods greatly improved the performance in metrics like peak signal-to-noise (PSNR) and structural similarity (SSIM).

\begin{figure}[t]
   \centering
   \includegraphics[width=1\linewidth]{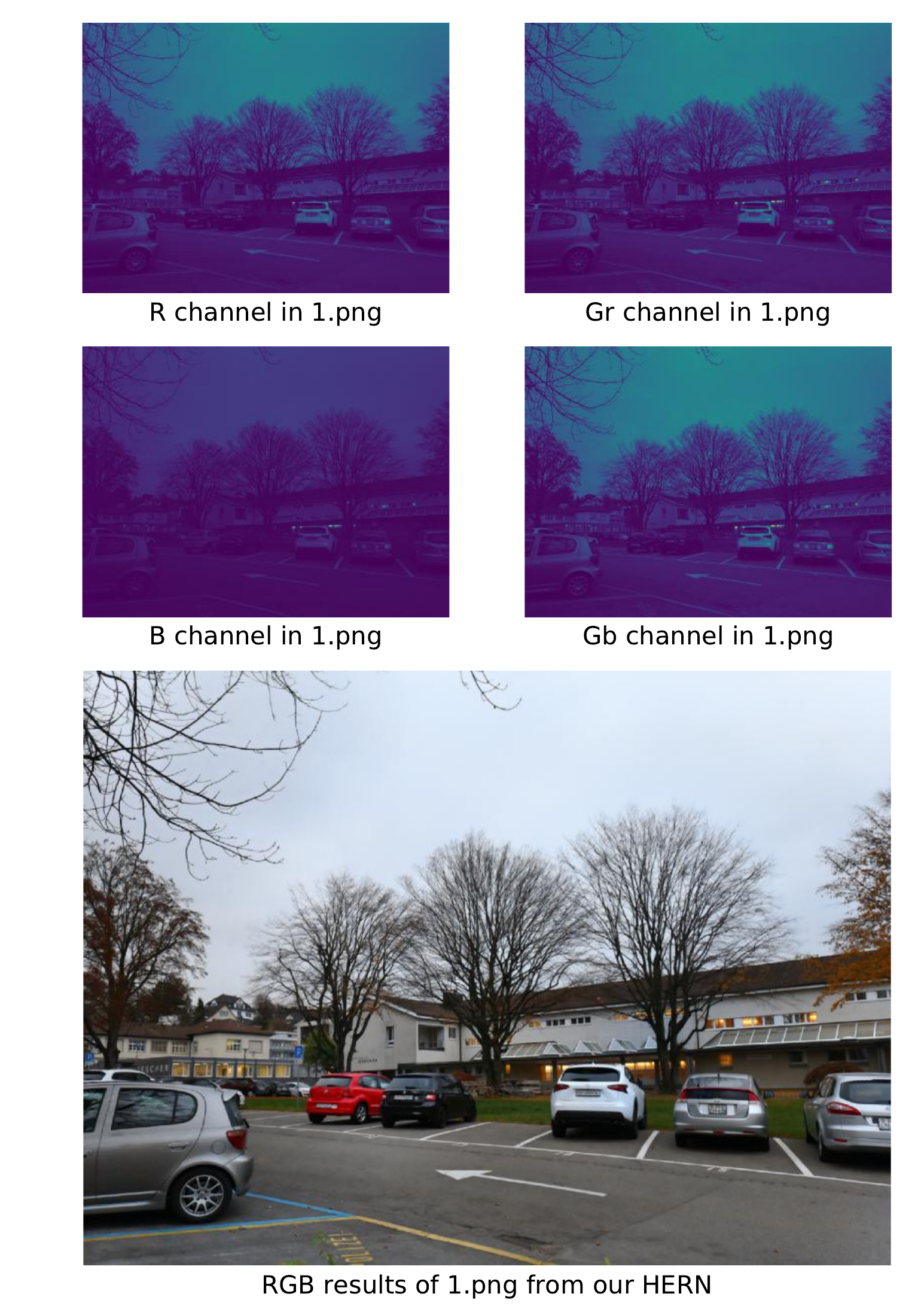}
   \caption{Visualization of each channel in the RAW image and its corresponding RGB results reconstructed by HERN.}
   \label{fig:raw_rgb}
\end{figure}

To further improve the performance and visual quality, recently works~\cite{abdelhamed2019ntire, xu2019towards} tend to train networks on unprocessed images.
More specifically, the proposed networks are learning the mapping between the degraded RAW image with the desired RGB image.
This is because unprocessed RAW images could provide more information than processed RGB images, thus more additional information could be used to reconstruct more accurately enhanced images. 
However, the RAW images outputted from the image sensors with Bayer filters consist of four different channels which increases the difficulty of mapping learning.
Different from RGB channels, the four channels in RAW (RGBG) is red (R), green (G), blue (B), and alternating green (G).
We provide an example of RAW image and its corresponding restoration results of our HERN in Figure~\ref{fig:raw_rgb}.
It should be noticed that each channel in the RAW image is not the same as the [R, G, B] channel in the RGB image.
Therefore, it is important to perform image processing operations like color correction and tone mapping before learning the mapping function.
To fully utilize the additional information from RAW images, using global information~\cite{gharbi2017deep} is as crucial as using local information.
The simplest way is using higher-resolution image patches instead of widely used low-resolution image patches during training.
However, current state-of-the-art methods are limited to using the low-resolution image patches due to the huge GPU memory usage.
The Figure~\ref{fig:gpu_memory} shows that the GPU memory usage will increases when the resolution of image patches increase.
Therefore, it is unrealistic to train state-of-the-art methods on one GPU card with image patches of 224 * 224 pixels (the maximum resolutions of image patches in the datasets).

To solve aforementioned problems, we propose a HighEr-Resolution Network (HERN), which consists of a dual-path network and a pyramid full-image encoder.
Especially, The dual-path network consists of a multi-scale module for local feature extraction and a modified residual in residual module for global information learning.
In the local information path, we use Multi-Scale Residual Blocks (MSRBs~\cite{li2018multi}) to full exploit local information on different scales.
In the global information path, we use the modified residual in residual (RIR) module as our backbone.
The residual in residual (RIR) module was inspired by the Residual Channel Attention Network (RCAN\cite{zhang2018image}).
However, the introduced channel attention mechanism in RCAN will cost huge GPU memory.
Therefore, we remove all channel attention layers to reduce the GPU memory usage and speed up training process.
Furthermore, we add two additional convolutional layers and two deconvolutional layers before and after the residual in residual (RIR) module like the PFFNet~\cite{mei2018progressive}, respectively.
These layers downsample the full-resolution image features into quarter size before learning and upsampling these features into the original size after learning.
Thus, we can train the model in higher resolution input image patches to extract more global information while with less GPU memory usage.
In addition, we add a pyramid full-image encoder to extract full-image information on a fixed resolution.
This additional encoder enable the network to process results with arbitrary resolutions.
Finally, the extracted local and global image features are concentrated together for the final RGB image reconstruction.
In summary, our contributions are:

\begin{figure}[t]
   \centering
   \includegraphics[width=1\linewidth]{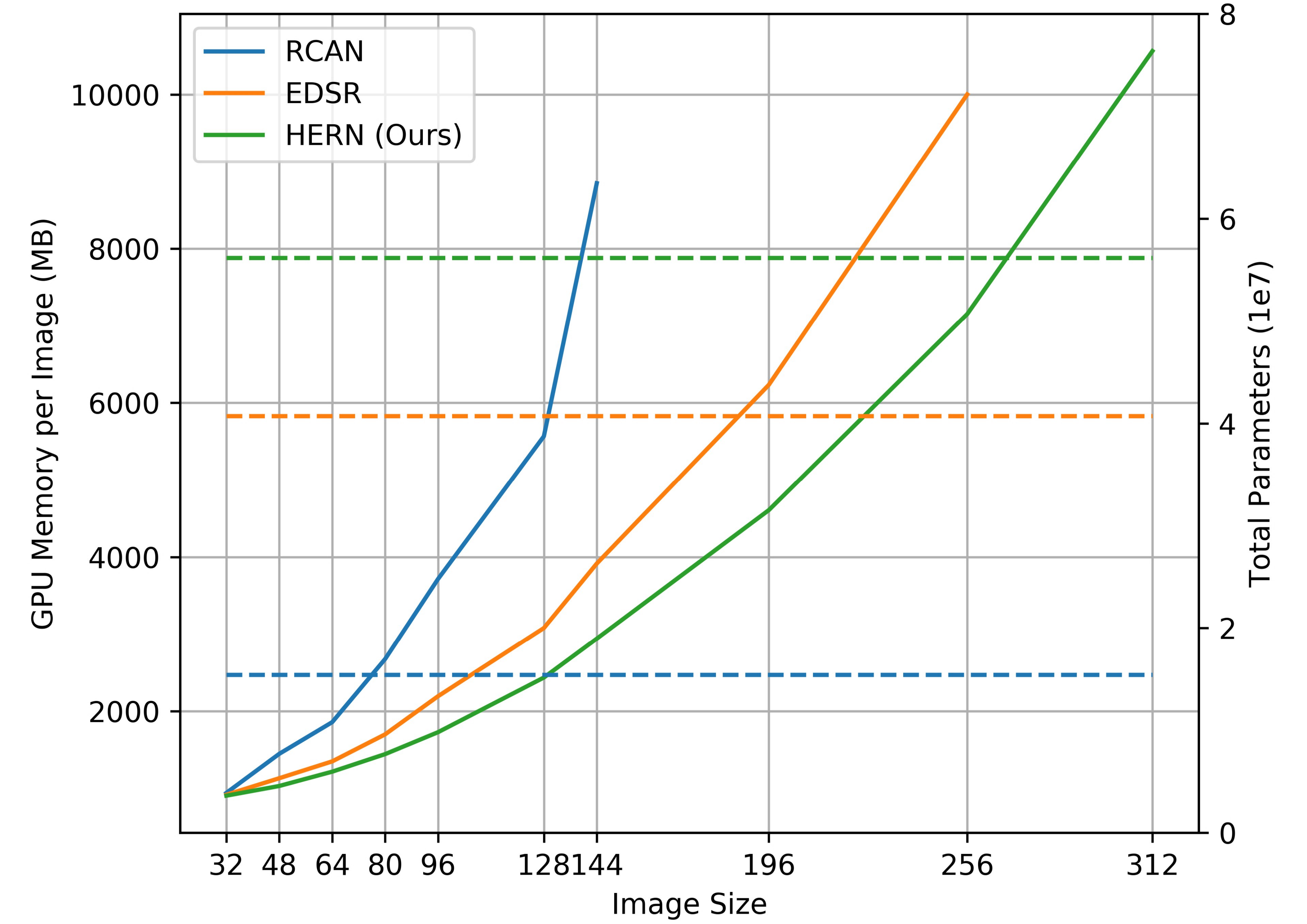}
   \caption{In this experience, we only use one  NVIDIA TitanX GPU and set the batch size to 1.
   Obviously, during training, GPU memory will increase as the resolution of the input image increases.
   The maximum input patch size that RCAN can handle is 144 * 144 while our proposed HERN can suitable for 312 * 312.}
   \label{fig:gpu_memory}
\end{figure}

(i). We propose a HighEr-Resolution Network (HERN), which can fully learning local and global information in high-resolution image patches. 

(ii). We propose a progressive training method to solve the instability issue and accelerate model convergence.

(iii). Our HERN won second place on track 1 (Fidelity) and won first place on track 2 (Perceptual) in the AIM2019 RAW to RGB Mapping Challenge.

\begin{figure*}
   \centering
   \includegraphics[width=1\linewidth]{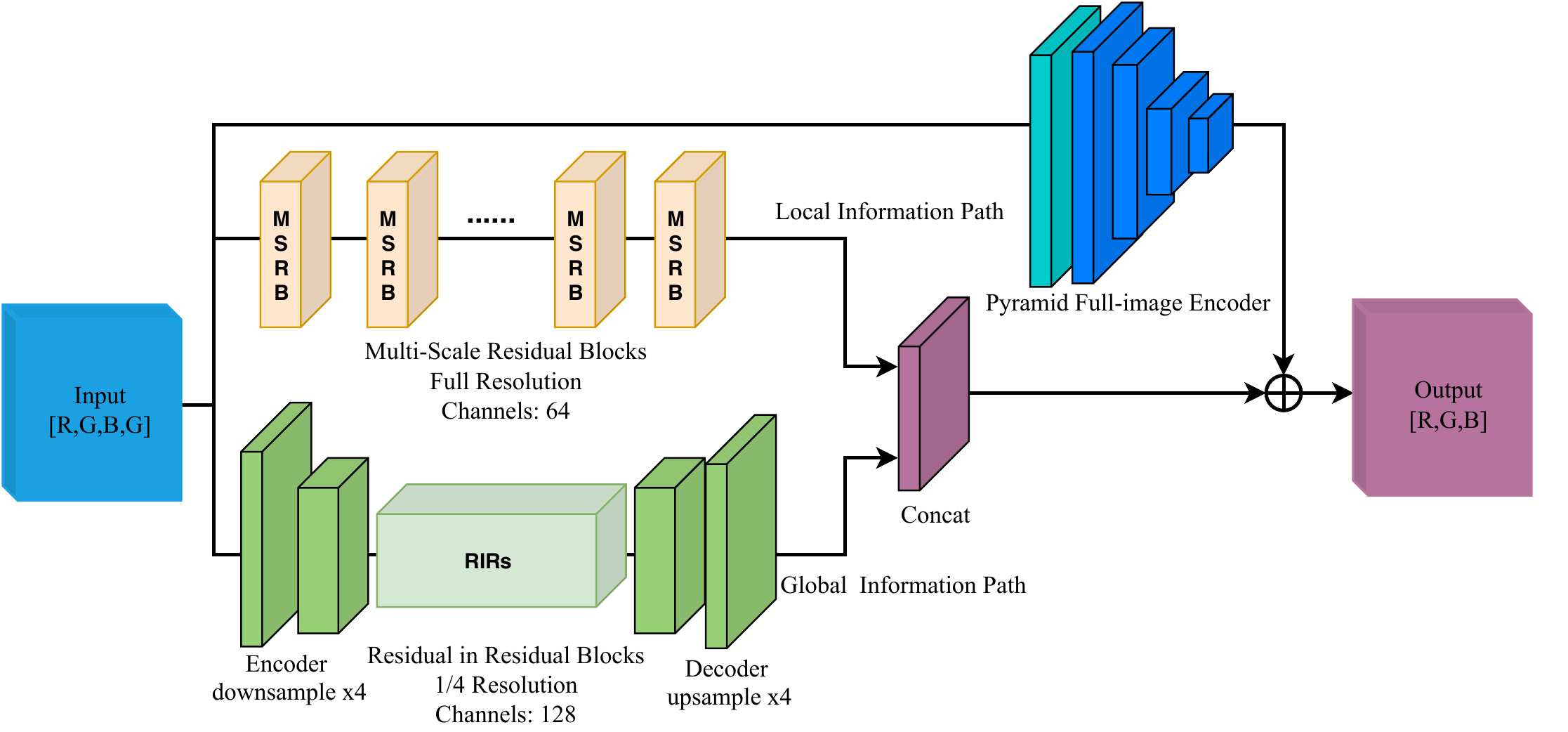}
   \caption{The architecture of our proposed HERN, which consists of a dual-path network and a pyramid full-image encoder.
   The pyramid full-image encoder (top) processes fix-resolution images $\mathrm{I}_{fix}$ generated by bilinear downsample, usually 192*192 pixels.
   The local information path (middle) processes full-resolution features $\mathrm{I}$ output by the first layer.
   The global information path (bottom) perform reconstruction on low-resolution features $\tilde{\mathrm{I}}$ output by the encoder layer, and reconstructed features are upsampled by the decode layers.}
   \label{fig:architecture_overview}
\end{figure*}

\section{Related Works}
\subsection{Image Enhancement}
Image enhancement aims to reconstruct the desired image from its degraded image.
Dong et al.~\cite{dong2015image} first proposed the CNN-based learning method (SRCNN) to directly learned the mapping between low-resolution images with high-resolution images, which achieves significant improvement than traditional learning-based methods.
Following this work, Cai et al.~\cite{cai2016dehazenet} and Ren et al.~\cite{ren2016single} use deeper CNNs to learn the mapping directly from haze-images to clear images.
These works proved that deeper networks with more parameters will achieve higher performance in terms of PSNR and SSIM.
Later work like EDSR~\cite{lim2017enhanced} and RCAN~\cite{zhang2018image} further use deep residual network up to 800 layers and achieve state-of-the-art performance.
Although the fidelity performance of these models is improved, the results tend to have indistinct details due~\cite{zhao2016loss} to the use of Mean Square Error (MSE) loss only.
To solve this problem, Ignatov et al.~\cite{ignatov2017dslr} replaced the MSE loss with a composite perceptual loss function, which got better perceptual results on translating ordinary photos into DSLR-quality images.
Different from ~\cite{ignatov2017dslr}, Gharbi et al.~\cite{gharbi2017deep} proposed the HDRNet to extract features in two different resolutions and bilateral slicing operations are used to reconstruct the low-resolution images into full resolution.
The HDRNet employs little parameters but can perform real-time enhancing and generate results in arbitrary resolutions with the same enhancing effects.
However, all the research mentioned above except for HDRNet assume the enhancing pattern is local and only low-resolution image patches are used for training.
Inspired the HDRNet~\cite{gharbi2017deep}, we aim to explore a more efficient model that can suitable for high-resolution input image patches.

\subsection{Image Demosaicing}
Image demosaicing include a set of operations, which are used to reconstruct color images from color filter arrays from an image sensor. 
The unprocessed RAW images contain the original information from the sensor and noise information from the lens or sensors.
So the demosaicing operation aims to remove the color artifacts like chromatic aberration, aliasing, and preserving the useful information as much as possible.

Recent works utilize the unprocessed information from RAW images to learn the mapping from RAW to RGB directly. 
For example, Cheng et al.~\cite{chen2018learning} developed a network operating on raw images directly without tradition image processing pipeline.
By employing the network on low-light image enhancing, they achieve more promising results than tradition image processing methods.
Brooks et al.~\cite{brooks2019unprocessing} introduce a technique to generate unprocessed images by inverting each step in demosaicing and then use the \emph{unprocessed} images to model noisy RAW images to RGB images by a simple CNNs.
The model achieves more accurate denoising results due to abundant information from the RAW images.
Following previous works, we aim to make full use of the abundant information in the RAW images to reconstruct high-quality RGB images in this work.

\section{HighEr-Resolution Network (HERN)}
In this section, we describe the proposed network architecture and corresponding training method.
We first discuss the details of each component in the network and then introduce the progressive training method, as well as the strategies comparison between training using high-resolution patches and low-resolution patches. 
The complete architecture of the proposed HERN is visualized in Figure~\ref{fig:architecture_overview}. 

\subsection{Dual-path Network}
The most important part of HERN is a dual-path network, which consists of a global information path and a local information path.

\subsubsection{Global Information Path}
\begin{figure}[t]
   \centering
   \includegraphics[width=0.98\linewidth]{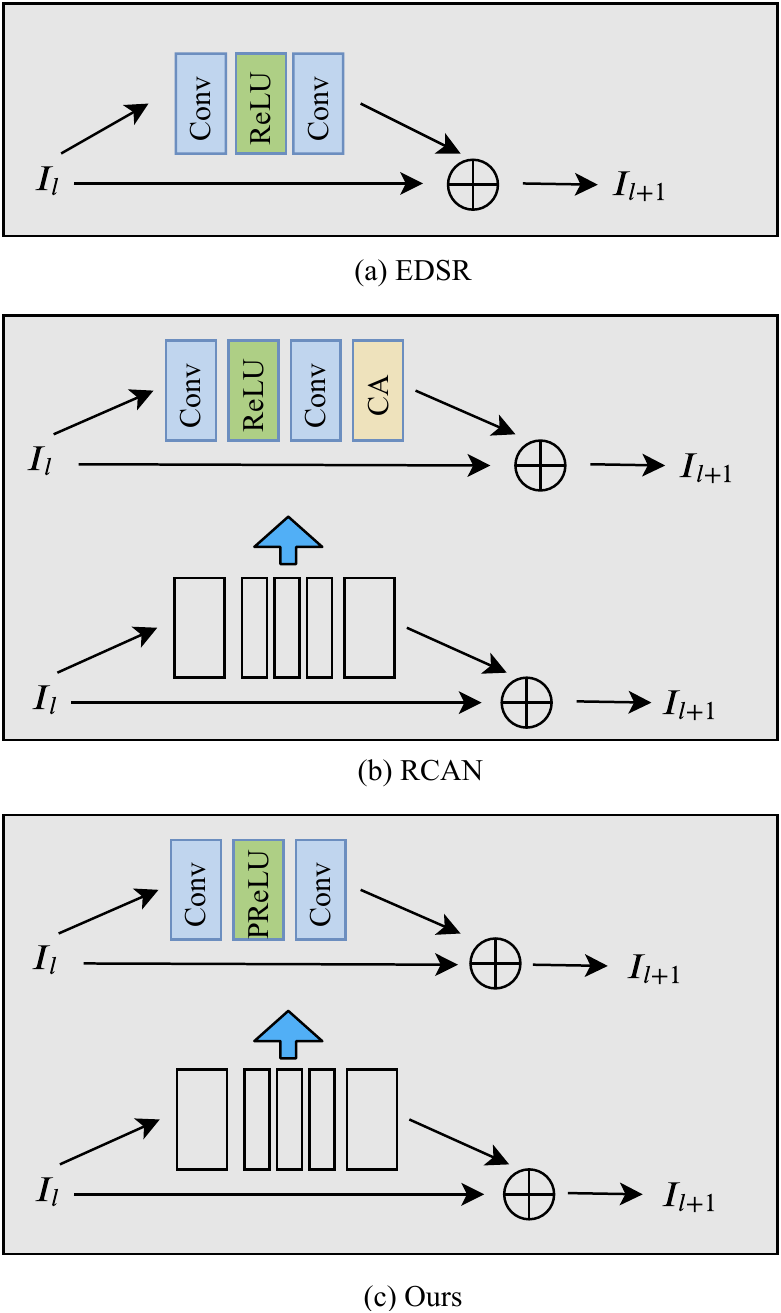}
   \caption{Architecture comparison between our modified RIR, the original RCAB in RCAN and the RB in EDSR.}
   \label{fig:rir}
\end{figure}

Residual in Residual (RIR)~\cite{zhang2018image} blocks have shown superior performance in the task of image super-resolution.
It consists of $G$ residual groups and each residual group consists of $B$ residual channel-attention blocks (RCABs).
By employing this architecture, it can be stacked into very deep networks without the vanishing gradient problem.
However, the proposed channel attention~\cite{hu2018squeeze} mechanism in each RCAB increase the memory usage greatly, especially when applied to high-resolution image features.
To fully exploit the global information, we remove all channel attention units in each block and stack the rest together with a global information path.
In addition, we replace the ReLU layer in each RIR block with the PReLU~\cite{he2015delving} layer since it provides penalties for negative values during training.
The difference between the original RCAB in RCAN and our propose RIR module are visualized in Figure~\ref{fig:rir}.

Besides, to further reduce the GPU memory usage, we introduce the Autoencoder mechanism into the global information path.
In other words, two convolutional layers with $stride=2$ (encoder) are applied before the RIRs module.
Therefore, the resolution of input image features are downsampled into a quarter of the original and the GPU memory usage is also reduced into a quarter of the original. 
After sufficient feature extraction by the RIRs, two deconvolutional layers with $stride=2$ (decoder) are applied in the tail of the path to upsampling these image features into the original size. 
This method can extract abundant global information, thus benefits for the final image reconstruction.

\subsubsection{Local Information Path}
\begin{figure}[ht]
   \centering
   \includegraphics[width=0.98\linewidth]{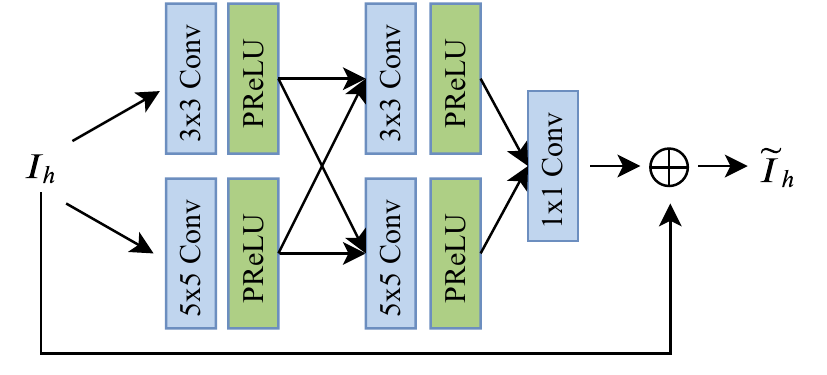}
   \caption{The architecture of MSRB in the local information path. We use PReLU layer in each MSRB instead of ReLU layer.}
   \label{fig:msrb}
\end{figure}

Although the introduced global information path has a large receptive field, it destroys local information such as textures and edges.
However, plenty of studies have shown that, image textures and edges are crucial in visual quality measurement. 
Therefore, we introduce a parallel local information path to recover the local information, which directly processes the feature without downsampling.
As shown in the Figure~\ref{fig:architecture_overview}(middle), the local information path consists of $M$ Multi-scale Residual Blocks (MSRBs~\cite{li2018multi}).
The MSRB use two convolutional layers with different kernel sizes (3x3 and 5x5) to allows the network to extract image features in different scales and combines different image features through the feature exchange mechanism.
Moreover, to fully utilize image features at difference scale, we add one bottleneck layer with 1x1 kernel in the tail of the block to fuse theses feature and reduce the GPU memory usage.
We visualize the architecture of the MSRB in Figure~\ref{fig:msrb},
where $I_h$ is the full-resolution image features and $\tilde I_h$ is the processed full-resolution image features.

\subsection{Pyramid Full-Image Encoder}
\begin{figure}[t]
   \centering
   \includegraphics[width=1\linewidth]{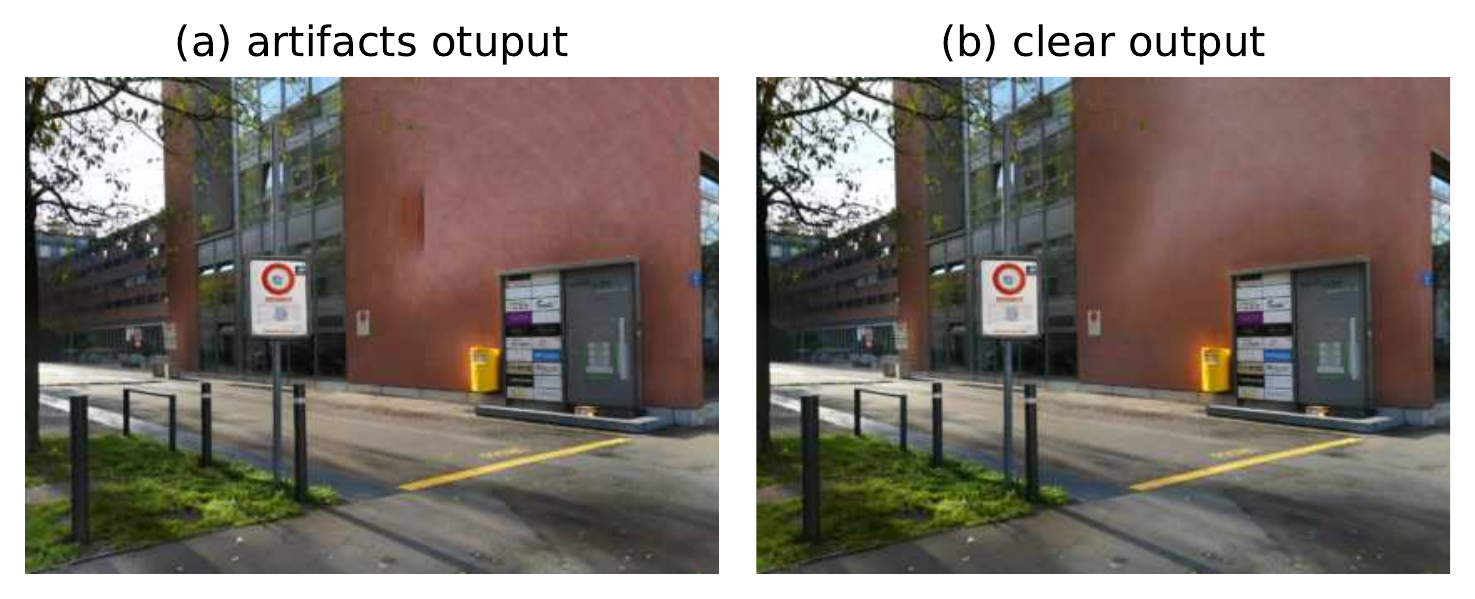}
   \caption{
   (a) shows the local artifacts in the wall area, which is generated by the model without the Pyramid Full-Image Encoder. 
   (b) shows clear results in the wall area, which is generated by the model with the Pyramid Full-Image Encoder.}
   \label{fig:artifacts}
\end{figure}

Different from the global information path proposed above, the pyramid full-image encoder produces a $n$-dimensional vector, which is then added to the fused features produced by the global information path and the local information path.
In this module, the input full-images are firstly processed by $P$ stacked convolutional layers with $stride=2$ and then one adaptive average pooling is adopted to reduce the dimension of features into 1*1. 
By doing this, the encoder encodes the whole images into high-level characteristics.
Due to the awareness of the global receptive field, it performs regularization on the output images, which improves the reconstruction performance, especially in the high-resolution testing.
We show one example of the local artifacts in the flat area generated by the networks in Figure~\ref{fig:artifacts}.
Obviously, the model with the Pyramid Full-Image Encoder can reconstruct more accurate results.
This because the introduced Pyramid Full-Image Encoder can extracts high-level characteristics of the input image and the extracted high-level characteristics can regularize local artifacts, especially in the flat area.

\subsection{Progressive Training} \label{sec:progressive}
Training with high-resolution images is not the same as training with low-resolution images, due to the huge GPU memory usage and the slow convergence speed.
Recently, many studies have been proposed to solve this problem.
For example, Karras et al.~\cite{karras2018progressive} processed a Progressive Growing GAN (ProGAN) to generate high-resolution images.
ProGAN grows the layers of generator and discriminator progressively, as well as growing the resolution of input images.
The results show that this training technology can speed up and stabilize the training process.
However, this techniques is not suitable for supervised image enhancement tasks because these model require many parameters at the initial training stage.
Directly using the training techniques of ProGAN, the model with fewer layers get bad parameters due to overfitting or mode collapse.

In this work, instead of growing layers and resolution of inputs progressively, we keep the same architecture of networks all the time and only the resolution of the input image is gradually increased.
Furthermore, without new parameters added during changing resolutions, the proposed network quickly adaptive to the new resolution inputs.
By training the network with low-resolution images, the network converges more quickly than directly training with high-resolution images.
The convergent network can then be adapted to the new network that training with higher-resolution images.
In this way, we can reduce the whole training time since most of the training process is in the lower-resolution level.

\section{Experiments and Results}
In this section, we describes the datasets used for training and validation, the training details, and the ensemble strategy.
Furthermore, fidelity comparison on image patches and visual quality comparison on full-resolution images are also presented.

In the global information path, we set $G$=16, $B$=10, and each layer have 128 filters.
In the local information path, we use 8 MSRBs for multi-scale image features extraction and each layer 64 filters.
Meanwhile, the Pyramid Full-Image Encoder contains one bilinear interpolation layer and 6 convolutional layers.

\begin{figure*}[htbp]
   \centering
   \captionsetup{justification=centering}
   \captionsetup[subfigure]{labelformat=empty}
   \begin{subfigure}[b]{0.18\linewidth}
      \centering
      \includegraphics[width=1\linewidth]{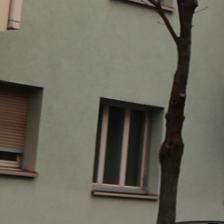} 
      \caption{PSNR / SSIM}
   \end{subfigure}
   \begin{subfigure}[b]{0.18\linewidth}
      \centering
      \includegraphics[width=1\linewidth]{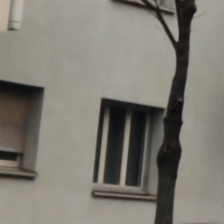} 
      \caption{19.509 / 0.856}
   \end{subfigure}
   \begin{subfigure}[b]{0.18\linewidth}
      \centering
      \includegraphics[width=1\linewidth]{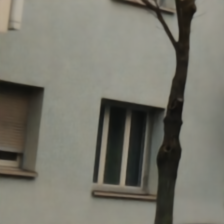} 
      \caption{21.798 / 0.867}
   \end{subfigure}
   \begin{subfigure}[b]{0.18\linewidth}
      \centering
      \includegraphics[width=1\linewidth]{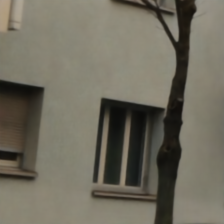}
      \caption{24.550 / \textbf{0.874}}
   \end{subfigure}
   \begin{subfigure}[b]{0.18\linewidth}
      \centering
      \includegraphics[width=1\linewidth]{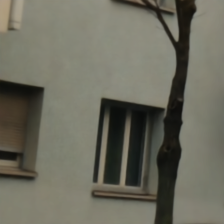}
      \caption{\textbf{24.857} / \textbf{0.874}}
   \end{subfigure}

   \begin{subfigure}[b]{0.18\linewidth}
      \centering
      \includegraphics[width=1\linewidth]{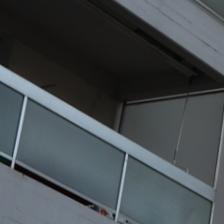} 
      \caption{PSNR / SSIM}
   \end{subfigure}
   \begin{subfigure}[b]{0.18\linewidth}
      \centering
      \includegraphics[width=1\linewidth]{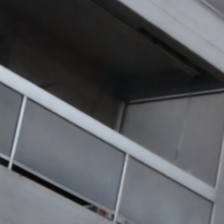} 
      \caption{21.166 / 0.857}
   \end{subfigure}
   \begin{subfigure}[b]{0.18\linewidth}
      \centering
      \includegraphics[width=1\linewidth]{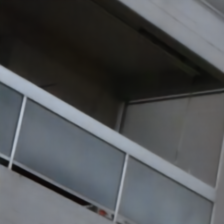} 
      \caption{24.560 / 0.875}
   \end{subfigure}
   \begin{subfigure}[b]{0.18\linewidth}
      \centering
      \includegraphics[width=1\linewidth]{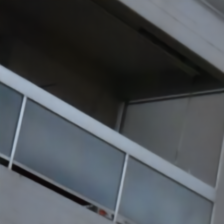}
      \caption{25.374 / 0.878}
   \end{subfigure}
   \begin{subfigure}[b]{0.18\linewidth}
      \centering
      \includegraphics[width=1\linewidth]{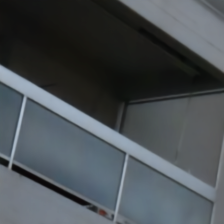}
      \caption{\textbf{25.577} / \textbf{0.879}}
   \end{subfigure}

   \begin{subfigure}[b]{0.18\linewidth}
      \centering
      \includegraphics[width=1\linewidth]{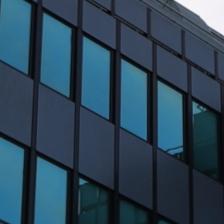} 
      \caption{PSNR / SSIM \\ Ground True}
   \end{subfigure}
   \begin{subfigure}[b]{0.18\linewidth}
      \centering
      \includegraphics[width=1\linewidth]{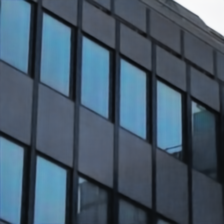} 
      \caption{20.032 / 0.82 \\ patch size = 72*72}
   \end{subfigure}
   \begin{subfigure}[b]{0.18\linewidth}
      \centering
      \includegraphics[width=1\linewidth]{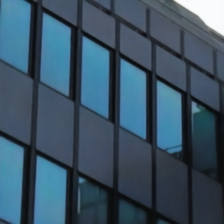} 
      \caption{23.803 / 0.875 \\ patch size = 144*144}
   \end{subfigure}
   \begin{subfigure}[b]{0.18\linewidth}
      \centering
      \includegraphics[width=1\linewidth]{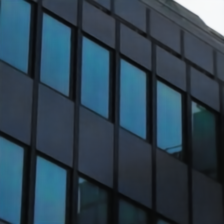}
      \caption{28.974 / 0.897 \\ patch size = 192*192}
   \end{subfigure}
   \begin{subfigure}[b]{0.18\linewidth}
      \centering
      \includegraphics[width=1\linewidth]{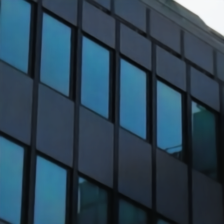}
      \caption{\textbf{30.349} / \textbf{0.901} \\ patch size = 224*224}
   \end{subfigure}
   \captionsetup{justification=justified}
   \caption{PSNR/SSIm and visual comparison of reconstructed images of models trained using different input patch sizes.}
   \label{fig:results}
\end{figure*}

\subsection{Datasets}
Zurich RAW2RGB Dataset~\cite{RAW2RGB2019report} (ZRR) is a novel datasets for the task of phone camera images enhancement.
It provides over 90K aligned image patch pairs include RAW images from a phone camera and RGB images from a DSLR correspondingly.
We divide the first 90000 image patch pairs used for training, and the rest 2139 image patch pairs used for fidelity validation.
Besides, 10 full-resolution RAW images without corresponding RGB images are used for perceptual validation.

\subsection{Training Details}
For training, we directly use RAW images as input and corresponding RGB images as ground truth.
Meanwhile, random horizontal and vertical flips are used for data augmentation.
According to ~\cite{zhao2016loss}, we use L1 loss instead of L2 loss to avoid getting stuck in a local minimum. L1 loss is denoted as follows:
$$
\mathcal{L}_1(\theta) = \frac{1}{N} \sum_{n=1}^{N} \left| \Phi (\rm{x}_i;\theta) - \rm{y}_i\right|,
$$
where $\Phi$ is the network function and $\theta$ represents the learnable parameters, $N$ is the batch size, and $\rm{x}_i$, $\rm{y}_i$ are the patch pairs of RAW image and RGB image.
As addressed in Section~\ref{sec:progressive}, we grow the resolution of inputs and ground truth progressively.
For the first 48 epoch, we use 72*72 pixels image patches and the learning rate is set to 1e-4.
For the later 36 epoch, we use 144*144 pixels image patches and the learning rate is set to 1e-5.
For the later 24 epoch, we use 192*192 pixels image patches and the same learning-rate.
For the final 8 epoch, we use 224*224 pixels image patches with constant learning-rate.
To fully utilize the GPU memory, the batch size is also decreasing as [16, 4, 2, 2] on a single NVIDIA TitanX GPU.
We use the random crop on full-resolution image patches instead of interpolation when changing resolution.
In addition, the Adam~\cite{kingma2014adam} optimizer with setting \( \beta_1=0.9, \beta_2=0.999 \) are used for all epochs.

\subsection{Ensemble Strategy}

Different from the data augmentation strategy used for training, ensemble strategy is used for testing.
Recently, plenty of studies have shown that the introduced ensemble strategy can further improve the model performance and robustness.
Therefore, we introduce the self-ensemble strategy and the epoch-ensemble strategy for the fidelity validation, and the perceptual validation only uses self-ensemble.

\begin{table}[htbp]
   \centering
   \resizebox{0.45\textwidth}{!}{ %
  \begin{tabular}{l|cc|c}
     \toprule
     Epoch & Self-Ensemble & Epoch-Ensemble & PSNR (dB) / SSIM \\ 
     \midrule
     \#113 & & &  23.105 / 0.810  \\
     \#113 & \checkmark & & 23.201 / 0.818  \\
     \#115 & & & 23.162 / 0.811   \\
     \#115 & \checkmark & & 23.254 / 0.818  \\
     \#113 \& \#115 & \checkmark & \checkmark & 23.304 / 0.818  \\
     \bottomrule
   \end{tabular}}%
   \caption{Ablation study on using different ensemble strategy during testing.
   The self-ensemble strategies increases performance in PSNR and SSIM greatly.
   The epoch-ensemble increases performance in PSNR only.}
   \label{tab:ablation}
\end{table}

The self-ensemble operation generates 3 different images by using the horizontal flip, the vertical flip, and the horizontal-vertical flip.
4 different images including the original image are input into the same network and 4 different outputs are flipped by inverse operations.
For the final reconstructed image is the average results of these 4 outputs.
The epoch-ensemble averages output from the same network with different training epochs, especially some epochs that achieve the highest performance in the validation set.
We also found this strategy is useful when the network is overfitting in some patterns.
In Table~\ref{tab:ablation}, we show the results of the validation dataset with or without these strategies.
Obviously, the introduced self-ensemble and epoch-ensemble strategies show superior performance.

\subsection{Quantitative and Qualitative Evaluation}
We use the ensembled results from the epoch \#114 and \#115 in the quantitative validation, and epoch \#115 only for the qualitative evaluation.
Besides, the ensembled results are trained with 224*224 pixels image patches.
We compared the results from our final model with the results from the same model but using input patches of different resolution for training.
The comparison results are shown in Figure~\ref{fig:results}.
We can see clearly that using higher-resolution patches for training can achieve more accurate reconstruction results than low-resolution patches, especially in global exposure adjusting and color reconstruction.

\subsection{AIM2019 RAW to RGB Mapping Challenge}
This work was proposed for participating the AIM2019 RAW to RGB Mapping Challenge~\cite{RAW2RGB2019report}.
The challenge consists of two tracks: track 1 focus on obtaining the highest pixel fidelity to the ground truth, which measured by PSNR and SSIM.
The track 2 focus on achieving the best perceptual quality similar to the ground truth, which measured by Mean Opinion Score (MOS).
It should be noted that the rating of perceptual quality is scaled from 0 to 4 and 0 is the best.

During this competition, we use the same model for both track 1 and track 2.
Furthermore,  our model achieves second place in track 1 and first place in track 2, the results are shown in Table~\ref{tab:challenge}.
This fully demonstrates that our model achieves superior performance on images of arbitrary resolution and can improve the visual quality of full-resolution images.

\begin{table}[htbp]
   \centering
   \resizebox{0.45\textwidth}{!}{ %
   \begin{tabular}{c|c|c|c|c}
     \toprule
     Track & Rank & Team & Metrics & Performance \\
     \midrule
     Track 1 & 1 & $1^{st}$ & PSNR / SSIM & \textbf{\textcolor{red}{22.59}} / \textbf{\textcolor{red}{0.81}} \\
             & \textbf{2} & \textbf{HERN} & PSNR / SSIM & \emph{\textcolor{blue}{22.24}} / \emph{\textcolor{blue}{0.80}} \\
             & 3 & $3^{st}$ & PSNR / SSIM & 21.94 / 0.79 \\
             & 4 & $4^{st}$ & PSNR / SSIM & 21.91 / 0.79 \\
             & 5 & $5^{st}$ & PSNR / SSIM & 20.85  / 0.77 \\
      \midrule
      Track 2 & 1 & \textbf{HERN} & MOS & \textbf{\textcolor{red}{1.24}} \\
              & 2 & $2^{st}$ & MOS & \emph{\textcolor{blue}{1.28}} \\
              & 3 & $3^{st}$ & MOS & 1.46 \\
              & 4 & $4^{st}$ & MOS & 1.56 \\
              & 5 & $5^{st}$ & MOS & 1.92 \\
     \bottomrule
   \end{tabular}}%
   \caption{AIM2019 RAW to RGB Mapping Challenge results for two tracks.
            We use red text to indicate the best performance and blue text to indicate the second best performance.}
   \label{tab:challenge}
\end{table}

\subsection{Limitations}
\begin{figure}[htbp]
   \centering
   \begin{subfigure}[b]{0.48\linewidth}
      \centering
      \includegraphics[width=0.98\linewidth]{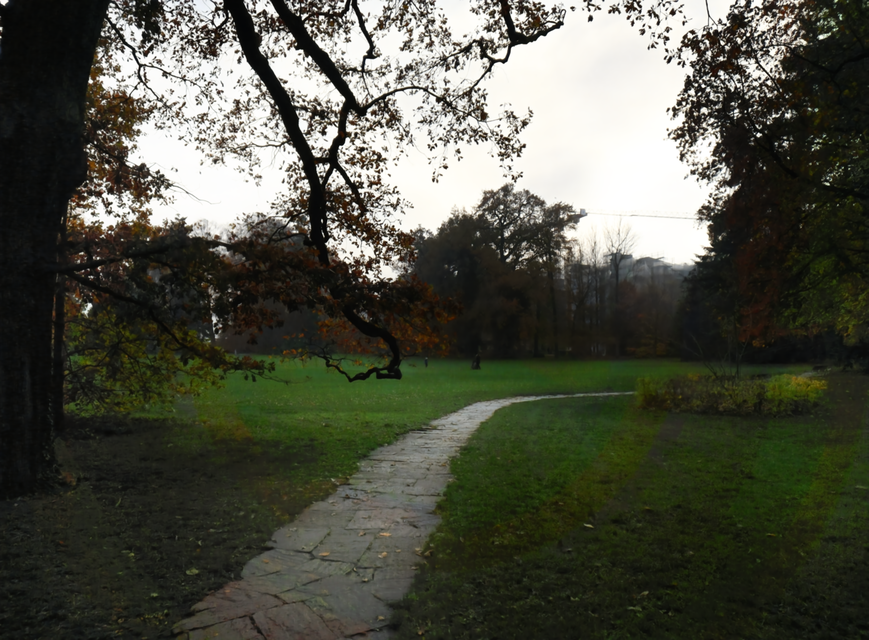} 
      \caption{results of 4.png}
   \end{subfigure}
   \begin{subfigure}[b]{0.48\linewidth}
      \includegraphics[width=0.98\linewidth]{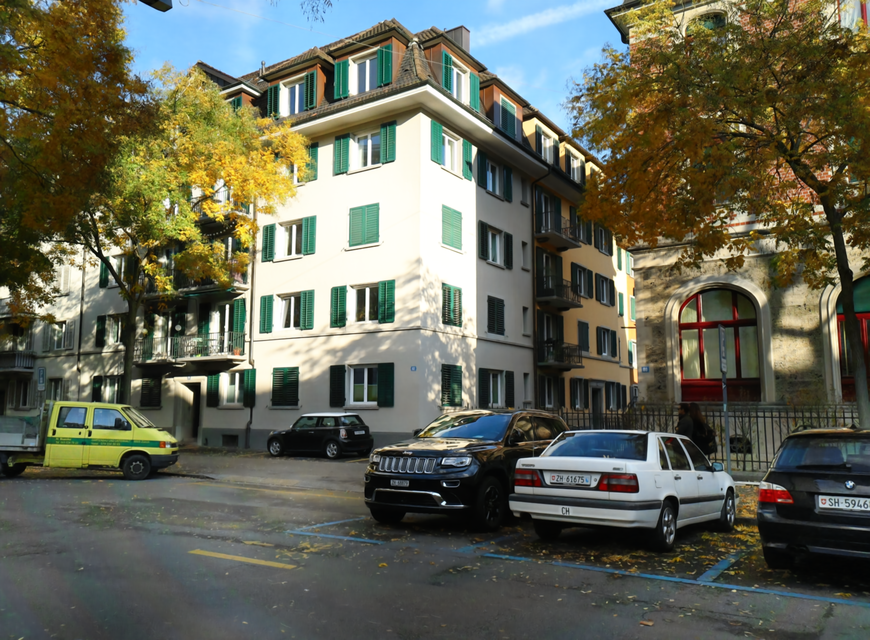}
      \caption{results of 5.png}
   \end{subfigure}
   \caption{Our HERN fails to generate correct results on these examples.
   In figure (a), the colorful circles occur in the bottom-right corner.
   In figure (b), the colorful circles occur in the bottom-left corner.}
   \label{fig:limitations}
\end{figure}

In Figure~\ref{fig:limitations} we show two examples of enhancing results in full-resolution.
There are some colorful bands around the center of the images.
These artifacts are similar to the vignetting of the camera lens and occur in both croped image patches and full-resolution images.
Since the provided RAW image is converted from 16-bit into 8-bit, it cannot contain such abundant colors, and faulty colors occur.
Our method cannot distinguish these color faults from enhancing effects, and these artifacts are unfortunately learned.
Thus, combining more accurate photography priors will be the focus of our future work.

\section{Conclusion}
State-of-the-art methods tend to modify network architectures for performance improvement. 
Different from them, this paper revises the main backbone from enhancement networks to adapt to training on high-resolution image patches.
For further utilizing benefits from using high-resolution patches, we propose a global information path, a local information path, and a pyramid full-image encoder in parallel.
By fusing three different outputs, the proposed HERN achieves the state-of-the-art performance on benchmark datasets.
Furthermore, it can process inputs of arbitrary resolution and generate similar effects on the same images without modifying networks architectures.
To stabilize the training process on high-resolution image patches, progressive training technology is proposed as well.
By progressively growing the resolution of inputs, it can get stability performance, as well as shortened training time.
The final implementation of this paper achieves second place in track 1 and first place in track 2 in the AIM2019 RAW to RGB Mapping Challenge.

{\small
\bibliographystyle{ieee_fullname}
\bibliography{egbib}
}

\end{document}